\documentclass[pra,twocolumn,showpacs,amsmath,amssymb,superscriptaddress,unsortedaddress,floatfix,longbibliography]{revtex4-1}
\usepackage{graphicx}
\usepackage{epstopdf}
\usepackage{latexsym}
\usepackage{amsmath}
\usepackage{graphics}
\usepackage{amssymb}
\usepackage{layout}
\usepackage{verbatim}
\usepackage{amsfonts,epsfig,dsfont}
\usepackage{natbib}
\usepackage{hyperref}
\usepackage{color}
\usepackage{physics}
\usepackage{listings}
\usepackage{mathtools}
\usepackage{bm}

	\newcommand{\rl}{\rangle\!\langle}
	
\begin{document}
		
		
		\title{Entanglement generation between a charge qubit and its bosonic environment during pure dephasing - dependence on environment size}
		
		\author{Tymoteusz Salamon}
		\affiliation{Department of Theoretical Physics, Faculty of Fundamental
			Problems of Technology, Wroc{\l}aw University of Science and Technology,
			50-370 Wroc{\l}aw, Poland}

		\author{Katarzyna Roszak}
		\affiliation{Department of Theoretical Physics, Faculty of Fundamental
			Problems of Technology, Wroc{\l}aw University of Science and Technology,
			50-370 Wroc{\l}aw, Poland}

		\date{\today}
		

		\begin{abstract}
			We study entanglement generated between a charge qubit and 
			a bosonic bath
			due to their joint evolution which leads to pure dephasing of the qubit. We tune the 
			parameters of the interaction, so that the decoherence is quantitatively
		independent of the number of bosonic modes taken into account and investigate,
			how the entanglement generated depends on the size of the
			environment. A second parameter of interest is the mixedness of the 
			initial state of the environment which is controlled by temperature.
			We show analytically that for a pure initial
			state of the environment, entanglement does not depend
			on environment size. For
			mixed initial states of the environment, the generated entanglement decreases with the increase of
			environment size. This effect is stronger for larger temperatures, when
			the environment is initially more mixed, but in the limit of an infinitely
			large environment, no entanglement is created at any finite temperature. 
			
		\end{abstract}

		\maketitle
		
		\section{Introduction}
		Decoherence due to the interaction of a qubit with its environment
		can often be modeled by classical noise with a good deal of accuracy.
		This is especially true with respect to pure dephasing, i.~e.~
		when decoherence does not disturb the occupations of the qubit, but 
		affects only its coherence (the off-diagonal elements of the density 
		matrix which are responsible for quantum behavior). Such loss of qubit
		coherence can always be mapped by a random unitary channel acting on 
		the system, where the channel describes the interaction with a \textit{fictitious}
		classical environment \cite{Helm_PRA09} (see Ref.~\cite{Crow_PRA14} for examples of constructions of such fictitious classical environments for different types of open quantum systems).
		The existence of such a mapping does not invalidate the importance of
		qubit-environment entanglement in pure-dephasing scenarios, since
		the presence of entanglement in the system will influence the evolution
		of the system in general, e.~g.~it changes the state of the environment
		pre- and post-measurement \cite{roszak15a}. Furthermore, if the study of a
		quantum system
		is not limited to its preparation, allowing it to evolve freely, 
		and then measurement, but also involves some manipulation of the system,
		such as performing gates \cite{calarco03,economou06,economou07}, coherence maximizing schemes \cite{viola98,viola03,roszak15b}, 	
		etc.~then the presence of system-environment entanglement 
		will influence the end result
		and can be higly relevant.
		
		Hence, it is at least in principle possible to have different qubit-environment
		setups, in which the pure dephasing of the qubit is qualitatively and 
		quantitatively the same, but the origin of the dephasing is different, since
		it can, but does not have to be the result of entanglement generation.
		If the whole system is always in a pure state
		(for an evolution described by a Hamiltionian this is equivalent to the
		initial states of the qubit and environment being pure)
		pure dephasing is unambiguously related to entanglement \cite{Zurek_RMP03,Hornberger_LNP09}. In the case of mixed states,
		the relation between qubit coherence and qubit-environment entanglement
		is much more ambiguous and although entanglement not accompanied by dephasing
		is not possible, dephasing without entanglement is \cite{Eisert_PRL02}.
		In fact, the latter situation is often realized in real systems, especially
		in the case of large environments, high temperatures, or noise
		resulting from e.~g.~fluctiating semi-classical fields \cite{merkulov02,coish2004,cywinski2009,cywinski2009b}. The distinction between
		entangling and non-entangling evolutions is not trivial in itself,
		since the non-entangling case is not limited to random unitary evolutions
		\cite{Helm_PRA09,Helm_PRA10,novotny10,LoFranco_PRA12,Liu_SR12,chruscinski13},
		and a straightforward criterion for the generation of qubit-environment
		entanglement during pure dephasing has only recently been found
		\cite{roszak15a}. 
		
		We study the amount of qubit-environment entanglement generated
		during the joint evolution of a charge qubit interacting with
		a bath of phonons as an example of a realistic system in which
		the qubit undergoes strictly pure dephasing which is always accompanied
		by the creation of entanglement (with the exception of only the infinite-temperature situation) \cite{roszak15a}. The system is particularly convenient,
		because not only can the decoherence-curves be reproduced using an arbitrary
		number of phonon modes (for short enough times and large enough temperatures), but the results can be obtained in a semi-analytical
		fashion, which simplifies changing the number of phonon modes
		and later interpretation of the results. Furthermore,
		we control the initial level of mixedness of the environment by setting
		the temperature of the phonon bath.
		
		The correlation build-up between a system and its bosonic environment have thus far 
		been studied in the context of its relation towards non-Markovian dynamics \cite{pernice12}, decoherence (especially for mixed initial qubit states) \cite{pernice11}, and 
		two qubit correlation decay \cite{maziero12,costa16}.
		Furthermore, studies of system-environment entanglement 
		for a boson system and a bosonic
		environment have also been reported \cite{eisert02,hilt09}.
		In these studies, the focus was on the relation between 
		the appearance of quantum correlations
		with a \textit{large} environment with other quantum features
		of open system dynamics. In this paper, the quantity of highest importance is the 
		size of the environment, which is vital for the amount of entanglement
		generated in the whole system.
		
		We find that the dependence of generated entanglement on the temperature
		shows monotonously decreasing behavior which is steep above a physically
		well motivated threshold temperature. The temperature dependence is non-trivial
		even above this temperature, and for reasonably low temperatures displays 
		exponential decay, while for high temperatures the dependence becomes 
		proportional to $1/T^3$. 
		On the other hand, the dependence on the number of phonon modes
		is much more complex. For pure states, 
		the amount of entanglement generated does not depend
		on the number of phonon modes at all. Yet for any mixed environmental state
		(finite temperature) there is a pronounced dependence on the number of 
		phonon modes (on the size of the environment) of $1/n$ character (where $n$
		is the number of modes). The maximum entanglement generated
		throughout the evolution always decreases when the environment becomes
		larger, even though the decoherence curve is unaffected in the studied
		scenario for high enough temperatures. Furthermore,
		this decrease is steeper when the temperature is higher
		(the initial state is more mixed), but when the number of phonon modes
		approaches infinity (the continuous case), all finite-temperature entanglement
		vanishes.
		
		The article is organized as follows. In Sec.~\ref{sec2} we introduce
		the system under study, the Hamiltonian describing this system,
		and the full qubit-environment evolution resulting from this Hamiltonian.
		We furthermore describe, how the respective strengths of the bosonic
		modes are determined, so that the decoherence at short times
		and large enough temperatures 
		is qualitatively and quantitatively the same independently of the
		size of the environment. Sec.~\ref{sec3} contains a brief description
		of Negativity, which is the entanglement measure which is later used
		to quantify qubit-environment entanglement. In Sec.~\ref{sec35} the dependence
		of the purity of the whole system (which is constant throughout the evolution) on the initial state of the environment is determined, especially on the 
		temperature and size of the environment. Sec.~\ref{sec4}
		contains the
		results pertaining to the dependence of the generated entanglement
		on temperature and consequently, on the degree of mixedness of the 
		initial state of the environment, as well as the results
		concerning the effect of environment size on entanglement, when
		the characteristics of the resulting qubit decoherence remain unchanged. The conclusions
		are given in Sec.~\ref{sec6}.

	\section{The system and the evolution \label{sec2}}

		The system under study consists of a charge qubit 
		interacting with phonons.
		The Hamiltonian of this system is
		\begin{equation}\label{ham}
		H=\epsilon|1\rl 1|+\sum_{\bm{k}}\hbar\omega_{\bm{k}}b_{\bm{k}}^{\dag}b_{\bm{k}}
		+|1\rl 1|\sum_{\bm{k}}
		(f_{\bm{k}}^{*}b_{\bm{k}}+f_{\bm{k}}b_{\bm{k}}^{\dag}),
		\end{equation}
		where the first term describes the energy of the qubit
		($\epsilon$ is the energy difference between the qubit states $|0\rangle$
		and $|1\rangle$
		in the absence of phonons), the second term 
		is the Hamiltonian of the free phonon subsystem and the 
		third term describes their interaction. Here, $\omega_{\bm{k}}$ is the
		frequency of the phonon mode with the wave vector $\bm{k}$ 
		and 
		$b_{\bm{k}}^{\dag}$, $b_{\bm{k}}$ are 
		phonon creation and annihilation operators corresponding to mode $\bm{k}$
		and $f_{\bm{k}}$ are coupling constants. 
		
		The Hamiltonian (\ref{ham}) can be diagonalized exactly using the Weyl
		operator method (see Ref.~\onlinecite{roszak06b} for details; the same results
		can be obtained using a different approach \cite{paz-silva16a,paz-silva16b}). For a product initial state of the system
		and the environment, $\sigma(0)=|\psi\rangle\langle\psi |\otimes R(0)$,
		where the qubit state is pure, $|\psi\rangle = \alpha |0\rangle +\beta |1\rangle$,
		and the environment is at thermal equilibrium, 
		\begin{equation}
		\label{r0}
		R(0)=\frac{e^{-\frac{1}{k_B T} \sum_{\bm{k}}\hbar\omega_{\bm{k}}b_{\bm{k}}^{\dag}b_{\bm{k}}}}{\mathrm{Tr} 
		\left[ e^{-\frac{1}{k_B T} \sum_{\bm{k}}\hbar\omega_{\bm{k}}b_{\bm{k}}^{\dag}b_{\bm{k}}}\right]},
		\end{equation}
		where $k_B$ is the Bolzmann constant and $T$ is the temperature,
		the joint qubit-environment density matrix evolves according to
		\begin{equation}
		\label{mac1}
		\hat{\sigma}(t)=\left(
		\begin{array}{cc}
		|\alpha|^2\hat{R}(0)&\alpha\beta^*e^{-i\epsilon t/\hbar}\hat{R}(0)\hat{u}^{\dagger}(t)\\
		\alpha^*\beta e^{i\epsilon t/\hbar}\hat{u}(t)\hat{R}(0)&|\beta|^2\hat{u}(t)\hat{R}(0)\hat{u}^{\dagger}(t)
		\end{array}\right).
		\end{equation}
		Here, the matrix is written in the basis of the qubit states 
		$|0\rangle$ and $|1\rangle$,
		while the degrees of freedom of the environment are contained in the 
		density matrix $\hat{R}(0)$ and time-evolution operators acting only on the
		environmnet, $\hat{u}(t)$. The evolution operators can be found following
		Ref.~[\onlinecite{roszak06b}], and are given by
		\begin{eqnarray}
		\nonumber
		\hat{u}(t)&=&
		\exp\left[\sum_{\bm{k}} 
		\left( \frac{f_{\bm{k}}}{\hbar\omega_{\bm{k}}}(1-e^{-i\omega_{\bm{k}} t})b_{\bm{k}}^{\dag}
		-\frac{f_{\bm{k}}^{*}}{\hbar\omega_{\bm{k}}}(1-e^{i\omega_{\bm{k}} t}) b_{\bm{k}} \right)\right]\\
		&&\times
		\label{u}
		\exp\left[i\sum_{\bm{k}} 
		\frac{|f_{\bm{k}}|^2}{(\hbar\omega_{\bm{k}})^2}\sin \omega_{\bm{k}} t\right].
		\end{eqnarray}
		
		In order to obtain the density matrix of the qubit alone, a trace over the
		degrees of freedom of the environment needs to be performed,
		$\hat{\rho}(t)=\Tr_E \hat{\sigma}$. This yields a density matrix with 
		time-independent occupations and coherences which undergo decay governed by the function
		\begin{equation}
		\label {srednia}
		|\langle \hat{u}(t)\rangle| = \exp\left(-\sum_{\bm{k}}\abs{\frac{f_{\bm{k}}}{\hbar\omega_{\bm{k}}}}^2(1-\cos{\omega_{\bm{k}}t})(2n_{\bm{k}}+1)\right),
		\end{equation}
		where $n_{\bm{k}}={1}/({e^{\hbar\omega_{\bm{k}}/k_BT}-1)}$ is the Bose-Einstein
		distribution.
		
		If, as in our case, the quantity of interest is qubit-environment entanglement
		and not just the coherence of the qubit,
		the time-evolution of the full system density matrix (\ref{mac1})
		is needed. This can be found by acting with the evolution operator
		given by eq.~(\ref{u}) on the initial density matrix
		of the environment. The density matrix of the whole system $\hat{\sigma}$
		can be divided into four parts with respect to the way that the evolution
		operator acts on the density matrix of the environment, which correspond
		to the $|0\rangle$ qubit state occupation (for which the environment 
		remains unaffected $\hat{R}_{00}(t)=\hat{R}(0)$), the $|1\rangle$ qubit state occupation (for which 
		$\hat{R}_{11}(t)=\hat{u}(t)\hat{R}(0)\hat{u}^{\dagger}(t)$),
		and the two qubit coherences (with 
		$\hat{R}_{01}(t)=\hat{R}(0)\hat{u}^{\dagger}(t)$ when the qubit density matrix
		element corresponds to $|0\rangle\langle 1|$ and
		$\hat{R}_{10}(t)=\hat{u}(t)\hat{R}(0)$ when the qubit density matrix
		element corresponds to $|1\rangle\langle 0|$).
		
		Since the initial density matrix of the environment is 
		a product of density matrices for each boson mode $\hat{R}(0)=\bigotimes_{\bm{k}}\hat{R}^{\bm{k}}(0)$ and so is the
		evolution operator at all times $\hat{u}(t)=\bigotimes_{\bm{k}}\hat{u}^{\bm{k}}(t)$, each matrix 
		$\hat{R}_{ij}(t)$ ($i,j=0,1$) also has product form with respect to the different
		boson modes at all times, so each boson mode can be treated separately.
		This is convenient, since every $\hat{R}_{ij}^{\bm{k}}(t)$ matrix
		is in principle of infinite dimension (the number of phonons in each mode
		can be arbitrarily large; the actual distribution of states for a single phonon
		mode
		is governed by the temperature and the qubit-phonon coupling)
		and a reasonable cut-off needs to be implemented to keep the density matrix
		$\sigma$ manageable without the loss of physical meaning.  
		Note, that although the $\hat{R}_{ii}(t)$ matrices
		corresponding to the diagonal elements of the qubit density matrix
		are density matrices themselves, this is not always true for the $\hat{R}_{ij}(t)$ matrices with $i\neq j$ (which is a first indicator
		of qubit-environment entanglement).
		
		It can be shown that the evolution of any state of $m$ phonons in mode $\bm{k}$ is given by
		\begin{eqnarray}
		\label{lag}
		|m(t)\rangle_{\bm{k}}&=&\hat{u}^{\bm{k}}(t)|m\rangle_{\bm{k}}\\
		\nonumber
		&=&\sum_{p=-m}^{\infty}\left(
		\frac{f_{\bm{k}}}{\hbar\omega_{\bm{k}}}
		\right)^p\sqrt{\frac{m!}{(m+p)!}}\\
		\nonumber
		&&\times L_{m}^{(p)}
		\left(\left|\frac{f_{\bm{k}}}{\hbar\omega_{\bm{k}}}\right|^2\right)
		|m+p\rangle_{\bm{k}},
		\end{eqnarray}
		where $L_m^{(p)}(x)$ is a generalized Laguerre polynomial. Given the initial
		state of the environment, eq.~(\ref{lag}) is sufficient to find the time
		evolution of the whole system-environment density matrix $\hat{\sigma}$,
		since
		\begin{eqnarray}
		\nonumber
		\hat{R}_{11}(t)&=&\hat{u}(t)\hat{R}(0)\hat{u}^{\dagger}(t)=\bigotimes_{\bm{k}} \left(\sum_{m_{\bm{k}}=0}^{\infty}
		c_{m_{\bm{k}}}|m(t)\rangle_{\bm{k}\ \bm{k}}
		\langle m(t)|
		\right),\\
		\nonumber
		\hat{R}_{10}(t)&=&\hat{u}(t)\hat{R}(0)=
		\bigotimes_{\bm{k}} \left(\sum_{m_{\bm{k}}=0}^{\infty}
		c_{m_{\bm{k}}}|m(t)\rangle_{\bm{k}\ \bm{k}}
		\langle m|\right),\\
		\nonumber
		\hat{R}_{01}(t)&=&\hat{R}(0)\hat{u}^{\dagger}(t)=\hat{R}^{\dagger}_{10}(t).
		\end{eqnarray}
		Here, the initial occupations of each state $|m\rangle_{\bm{k}}$
		are found for a given temperature using eq.~(\ref{r0}),
		\begin{equation}
		c_{m_{\bm{k}}}=e^{-\frac{\hbar\omega_{\bm{k}}}{k_BT}m_{\bm{k}}}(1-e^{-\frac{\hbar\omega_{\bm{k}}}{k_BT}}).
		\end{equation}
		
		\subsection{Excitonic quantum dot qubits \label{secIIA}}
		The exciton-phonon interaction constants used in the calculations
		correspond to excitonic qubits confined in quantum dots
		\cite{roszak06b,krummheuer02,vagov03,vagov04}, where
		qubit state $|0\rangle$ corresponds to an empty dot,
		while state $|1\rangle$ denotes an exciton in its ground state
		confined in the dot. They 
		are given by
		\begin{equation}
		\label{cpl}
		f_{\bm{k}}
		=( \sigma_{\mathrm{e}} -\sigma_{\mathrm{h}} )
		\sqrt{\frac{\hbar k}{2\varrho V_{\mathrm{N}} c}}
		\int_{-\infty}^{\infty} d^3\bm{r}\psi^*(\bm{r})
		e^{-i\bm{k}\cdot\mathrm{r}}\psi(\bm{r}),
		\end{equation}
		describing the deformation potential coupling, which is the dominating 
		decoherence mechanism for excitons \cite{krummheuer02}.
		Hence, $\omega_{\bm{k}}=ck$, where $c$ is the speed of longitudinal sound
		and the phonon-bath is super-Ohmic.
		Here $\varrho$ is the crystal density, $V_{\mathrm{N}}$ unit cell
		volume, and
		$\sigma_{\mathrm{e,h}}$ are deformation potential constants for
		electrons and holes respectively.
		The exciton wave function $\psi(\bm{r})$
		is modeled as a product of two identical single-particle wave
		functions $\psi(\bm{r}_{\mathrm{e}})$ and $\psi(\bm{r}_{\mathrm{h}})$,
		corresponding to the electron and hole, respectively.
		
		The parameters used in the calculations correspond to small self-assembled 
		InAs/GaAs quantum dots, which are additionally assumed to be isotropic 
		(for the sake of simplicity when limiting the number of phonon 
		modes and with little loss of realism, when the evolution of coherence
		is found). The single particle wave functions 
		$\psi(\bm{r})$ are modeled by
		Gaussians with $3$ nm width in all directions. 
		The deformation
		potential difference is $\sigma_{\mathrm{e}}-\sigma_{\mathrm{h}}=9.5$ eV,
		the crystal density is $\varrho=5300$ kg/m$^{3}$, and the speed of
		longitudinal sound is $c=5150$ m/s. The unit cell volume for GaAs is 
		$V_{\mathrm{N}}=0.18$ nm$^3$ (note, that this volume does not enter into the decoherence
		function, but is relevant, when individual elements of the system-environment
		density matrix needed to evaluate entanglement are found).

		\subsection{Discretization \label{secIIB}}
		
		Typically for realistic systems, the number of phonon modes is very large
		and the summation over $\bm{k}$ can be substituted by integration, which 
		in spherical coordinates yields
		\begin{equation}
		\sum_{\bm{k}}\to\frac{V}{(2\pi)^3}\int_0^{2\pi}d\phi\int_0^{\pi}\sin\theta d\theta \int_0^{\infty}k^2dk.
		\end{equation}
		If the studied system has spherical symmetry, as do the quantum dots, the parameters of which are used in the calculations, the integration over the 
		angles can be performed
		analytically. In the following, when we study qubit-environment entanglement 
		and qubit decoherence due to the interaction with an environment which supports
		only a limited number of phonon modes, we do not differentiate the modes
		with respect to their direction, only with respect to the length of the wave
		vector. This means that we consider a simplified scenario, where phonon
		modes are averaged over all directions. 
		\begin{figure}[th]
			\begin{center}
				\unitlength 1mm
				\begin{picture}(75,55)(5,5)
				\put(0,0){\resizebox{85mm}{!}{\includegraphics{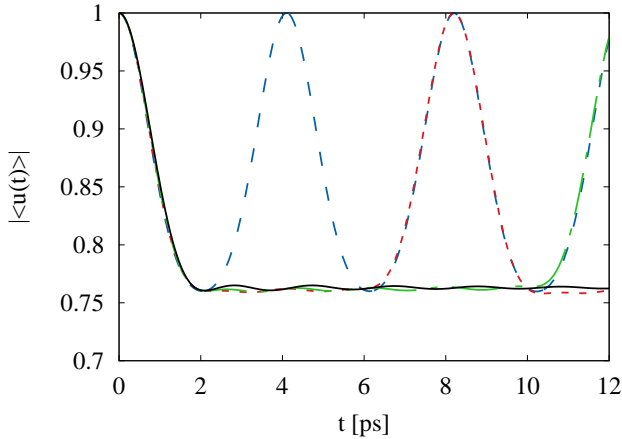}}}
				\end{picture}
			\end{center}
			\caption{\label{fig2} Evolution of the degree of qubit coherence at $T=6$ K for different numbers of phonon modes: $n=3$ - dashed blue line,
				$n=5$ - dotted red line, $n=7$ - dashed-dotted green line,
				$n=100$ - solid black line,}
		\end{figure}
		
		The actual discretization of the phonon modes is done
		only on the level of the length of the wave vector $k$.
		A minimum and maximum wave vector length is, somewhat arbitrarily, chosen,
		so that a large enough range of $k$ is considered to 
		account for different phonon modes with values of
		the function $|f_{\bm{k}}/\hbar\omega_{\bm{k}}|^2$ which are large enough
		to be relevant for both pure dephasing and entanglement generation.
		In the following they are always set to $k_{\mathrm{min}}=0.001$ nm$^{-1}$ and 
		$k_{\mathrm{max}}=0.9$ nm$^{-1}$.
		For a given number of phonon modes $n$, the range $[k_{\mathrm{min}},k_{\mathrm{max}}+k_{\mathrm{min}}]$ is evenly divided,
		and only wave vectors
		of lengths $k_i=(i-1)\Delta k+k_{\mathrm{min}}$, with 
		$\Delta k=k_{\mathrm{max}}/(n-1)$ and $i=1,2,...,n$, are taken into account.
		The slight off-set by of the range of $k$ by $k_{\mathrm{min}}$ allows
		for the decoherence of the qubit in the continuous case (when an infinite
		number of phonon modes is taken into account) to be well approximated 
		by only a few phonon modes for a wide range of temperatures as seen below.
		
		The decay of qubit coherence for a quantum dot interacting with an environment
		for which only a few discrete lengths of phonon wave vectors are allowed
		(in which only a few phonon modes are present) is plotted in Fig.~(\ref{fig2})
		for $T=6$ K (note that the coherence on the plot is limited from below
		by $0.7$ and not by the minimum value of the degree of coherence, $0$,
		for clarity). The initial drop of coherence which is present in the
		continuous case, and which is in the continuous case followed by
		a slight rise in coherence, after which the coherence 
		stabilizes at some finite value due to the super-Ohmic nature of the 
		environment
		(hence the term ``partial pure dephasing''; see Ref.~\cite{roszak06b}
		for details) is reproduced very well already when only three phonon modes are
		present in the environment. For larger boson mode numbers $n$, the longer-time
		features of decoherence are also reproduced for a finite time, 
		and the refocusing of the qubit
		(which is the result of the whole qubit-environment density matrix 
		returning to its initial state in the course of its unitary evolution)
		is further and further delayed in time with increasing number of phonon
		modes. 
		
		In fact, it is easy to reproduce continuous short-time behavior of the coherence
		even with small phonon mode numbers as long as the initial temperature is above
		some threshold, which depends on the phonon energy spectrum, and is here
		around $3$ K (this also obviously depends on the number of phonon modes and the 
		threshold is higher for higher mode numbers). Below this temperature, the approximation gets progressively worse, if the choice of phonon modes remains
		unchanged and spans the whole energy range where the spectral density is
		reasonably large.
		This is because the temperature enters into the calculation only through
		the initial state of the environment (which is at thermal equilibrium) 
		and, if only a few phonon modes
		with well separated energies are allowed, then below some temperature
		practically no phonons will be initially excited. This can be remedied by
		a redistribution of the phonon modes to lower energies, but since we wish
		to compare qubit-environment entanglement evolutions, when environments with
		different numbers of boson modes lead to the same dynamics of the qubit alone,
		but with a set system under study for a given number of modes,
		we simply restrict ourselves to higher temperatures. This means that for a set
		number o phonon modes the studied system is always qualitatively the same
		(and even though the initial state of the system depends on temperature,
		the types of phonons which can be excited remain unchanged).

	\section{Negativity \label{sec3}}
	
		The measure of entanglement which is most convenient (easiest to compute) 
		in the context
		of quantifying entanglement between a qubit (a small quantum system)
		and its environment (a large quantum system) is Negavitity \cite{vidal02,lee00a} (or equivalently logarithmic
		Negativity \cite{plenio05b}). The measure is based on the PPT criterion
		of separability \cite{Peres_PRL96,Horodecki_PLA96}, which does not detect
		bound entanglement \cite{Horodecki_PLA97,Horodecki_PRL98}. Fortunately,
		in the case of the evolution of a qubit initially in a pure state interacting
		with an arbitrary environment due to an interaction which can only lead to 
		pure dephasing of the qubit, bound entanglement is never formed
		\cite{horodecki00,roszak15a}, so non-zero Negativity is a good
		criterion for the presence of entanglement in the system and the value of 
		Negativity unambiguously indicates the amount of said entanglement.

		Negativity can be defined as the absolute sum of the negative eigenvalues
		of the density matrix of the whole system after a partial transposition 
		with respect to one of the two potentially entangled subsystems
		has been performed
		(it does not depend on which of the subsystems is chosen for the partial
		transposition),
		\begin{equation}
		N(\hat{\sigma})=\sum_i\frac{|\lambda_i|-\lambda_i}{2},
		\end{equation}
		where $\lambda_i$ are the eigenvalues of $\hat{\sigma}^{\Gamma_A}$, and
		$\Gamma_A$ denotes partial transposition with respect to system $A=Q,E$
		(qubit or environment). In the case of the studied system, it is particularly
		simple to perform partial transposition with respect to the qubit, as it is
		sufficient to exchange the off-diagonal terms in eq.~(\ref{mac1})
		to get the desired partially transposed state.    
		
		\section{Purity \label{sec35}}
		An important factor for the amount of entanglement generated between the
		qubit and the environment is the initial purity of the state of the whole
		system. Note, that since the qubit-environment evolution is unitary, the
		purity does not change with time, since
		\begin{eqnarray}
		\nonumber
		P(\hat{\sigma}(t))&=&\Tr\hat{\sigma}^2(t)=\Tr\left[
		\hat{U}(t)\hat{\sigma}(0)\hat{U}^{\dagger}(t)
		\hat{U}(t)\hat{\sigma}(0)\hat{U}^{\dagger}(t)
		\right]\\
		\nonumber
		&=&\Tr\hat{\sigma}^2(0)=P(\hat{\sigma}(0)).
		\end{eqnarray}
		
		Taking into account that 
		the initial state of the studied system is a product state
		and the state of the qubit subsystem is pure, we have  
		\begin{equation}
		P(\hat{\sigma}(0))=P(\hat{\rho}(0))P(\hat{R}(0))=P(\hat{R}(0)),
		\end{equation}
		so the purity of the system only depends on the initial purity of the
		density matrix of the environment.
		Furthermore, this initial density matrix is a product of
		the thermal-equilibrium density matrices for each phonon mode, so the purity
		is a product of the purities of the state of each mode, 
		$P(\hat{R}(0))=\prod_{\bm{k}}P(\hat{R}^{\bm{k}}(0))$. 
		The purity of the initial state of mode $\bm{k}$ is easily found from
		eq.~(\ref{r0}) and is given by
		\begin{equation}
		\label{prk}
		P(\hat{R}^{\bm{k}}(0))=
		\frac{\left(1-e^{-\frac{\hbar\omega_{\bm{k}}}{k_BT}}\right)^2}
		{1-e^{-2\frac{\hbar\omega_{\bm{k}}}{k_BT}}}.
		\end{equation}
		Since $e^{-\frac{\hbar\omega_{\bm{k}}}{k_BT}}$ tends to one with
		growing temperature more slowly
		than $e^{-2\frac{\hbar\omega_{\bm{k}}}{k_BT}}$, the numerator in eq.~(\ref{prk}) tends to zero much
		faster than the denominator, and the purity of the ininial state of a given 
		phonon mode is a decreasing function of temperature (which reaches
		zero for infinite temperature, since the dimension of the Hilbert space
		of each phonon mode is infinite). Consequently, the purity
		of the whole environment for a set choice and number of phonon modes is
		always a decreasing function of temperature as well.
		
		Less obviously, for the system under study the purity is also a decreasing function of the number of phonon modes. For a given number
		of modes $n$, $n$ wave vectors are taken into account which are evenly
		distributed throughout a set wave vector lengths $k$ where the 
		coupling constants are most relevant as explained in Sec.~\ref{secIIB},
		and the purity of their initial state is a product of the corresponding
		single-phonon-mode purities (\ref{prk}). The energy of each phonon mode
		is proportional to its wave vector length, $\omega_{\bm{k}}=ck$, 
		so the single-phonon-mode purity is an increasing function of $k$ for 
		any finite temperature. If the number of phonon modes is increased by one,
		each phonon mode is substituted by one with smaller wave vector length
		$k$
		(with the exception of the two phonon modes limiting the rangle of $k$), and hence, of lesser purity, and an additional phonon mode with a longer
		wave vector is taken into account (since now we are dealing with $n+1$ phonon modes evenly distributed
		over the same range). Consequently, the product of the new $n+1$ purities, which yields
		the purity of the initial state of the environment for an increased number
		of phonon modes, must be smaller for any finite temperature than the purity
		for $n$ modes. The exception is the zero-temperature case, for which 
		the purity of the initial environment is always equal to one, since the
		initial state is pure, and the infinite-temperature case, when the purity
		is always equal to zero. Hence, although for every number of phonon modes
		the purity is a decreasing function of temperature ranging from one to zero,
		the decrease is faster, if $n$ is larger.

\section{Results: Qubit-environment entanglement generation \label{sec4}}

\subsection{Temperature dependence \label{secVA}}

\begin{figure}[th]
			\begin{center}
				\unitlength 1mm
				\begin{picture}(75,55)(5,5)
				\put(0,0){\resizebox{85mm}{!}{\includegraphics{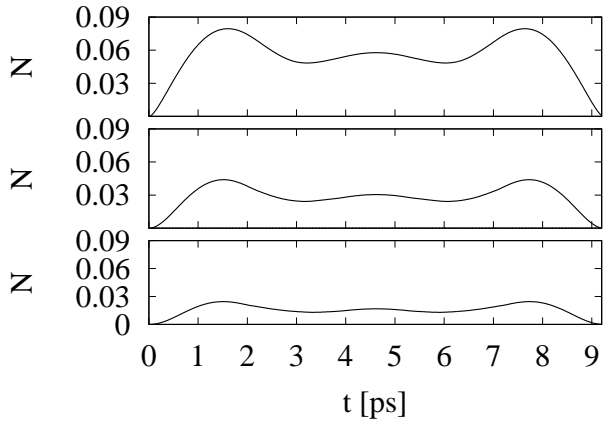}}}
				\end{picture}
			\end{center}
			\caption{\label{fig3} Entanglement evolution for $n=10$ boson modes at $T=6$ K
				(upper panel), $T=9$ K (middle panel), and $T=12$ K (lower panel).}
		\end{figure}

		The time-evolution of entanglement for the initial qubit state with
		$\alpha=\beta=1/\sqrt{2}$ (equal superposition) 
		quantified by Negativity is plotted
		in Fig.~(\ref{fig3}) for $n=10$ boson modes at three different temperatures
		(the temperature increases from top to bottom of the figure).
		The relatively large number of boson modes guarantees that the evolution
		of the qubit state
		due to the exciton-phonon interaction not only reproduces the fast
		initial	decay of qubit coherence which occurs in the first two picoseconds
		after the creation of the excitonic superposition
		for a continuous environment, but also gives a reasonable approximation
		of the coherence plateau for the next four picoseconds (up to slight oscillations
		which are absent when an infinite number of phonon modes is taken into account).
		The temperatures in the plot start at $6$ K, well above the threshold temperature, so phonon modes which are evenly
		distributed over the range of relevant coupling constants $f_{\bm{k}}$
		reproduce the dynamics of decoherence for short times well.
		The plots in  Fig.~(\ref{fig3}) capture a single cycle of qubit-environment
		evolution, so at the right end of the plots, the density matrix of the whole
		system returns to its initial state, and then the evolution is repeated
		(this is an unavoidable feature of systems with discrete spectra).
		
		\begin{figure}[th]
			\begin{center}
				\unitlength 1mm
				\begin{picture}(75,55)(5,5)
				\put(0,0){\resizebox{85mm}{!}{\includegraphics{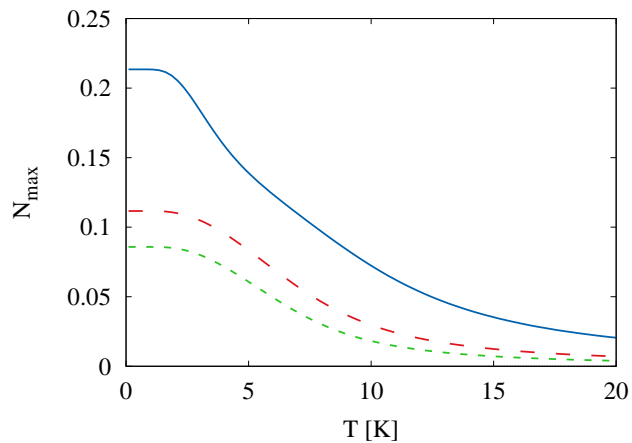}}}
				\end{picture}
			\end{center}
			\caption{\label{fig4} Maximum Negativity as a function of temperature for $n=2$
				(solid blue line), $n=4$ (dashed red line), and $n=6$ (dashed greeen line)
				wave vetors. }
		\end{figure}
		
		As can be seen, 
		changing the temperature does not change the qualitative features of
		the evolution of Negativity, but
		entanglement (at any given time) is a decreasing function
		of temperature. Contrarily, decoherence increases with temperature,
		so for higher temperatures the effect of the environment on the qubit 
		is larger (leading to stronger pure dephasing), but this is due to a buildup
		of classical qubit-environment correlations, since the amount of entanglement
		generated between the two subsystems decreases with decreasing purity of the 
		state.
		
		In Fig.~(\ref{fig4}) the dependence of the maximum Negativity reached
		during the pure dephasing evolution (Negativity reached at the first maximum
		which corresponds to the initial strong loss of coherence) is plotted as 
		a function of temperature for a choice of three different numbers of boson
		modes. Below around $T=2$ K, the maximum Negativity stabilizes at an
		almost fixed value. This is because the initial density matrix of the environment
		below this temperature becomes a very weak function of temperature, since
		only phonon modes with high energies compared to $k_BT$ are taken into account 
		(if there are only a few phonon modes allowed in the system). The density 
		matrix of the environment is then almost in the pure state 
		$\hat{R}(0)\approx |0\rangle\langle 0|$, 
		and the resulting qubit decoherence is
		no longer a good approximation of the continuous case.
		Note that in such situations, 
		the plateau in Negativity is strictly related to the discrete nature of the
		phonon energy spectrum. An agreement between continuous and few-phonon-mode
		decoherence could be reached also for low temperatures, but this would require
		changing $k_{\mathrm{max}}$ and would qualitatively change the system under 
		study (which we want to avoid). If the necessary 
		redistribution of
		the phonon modes taken into account (to account for decoherence well) were made,
		the plateau in low-temperature negativity would not be observed.
		
		At higher temperatures, maximum Negativity decreases strongly with temperature,
		regardless of the number of phonon modes taken into account, although the
		actual amount of entanglement in the system depends strongly on $n$. The shapes of the Negativity curves plotted in  
		Fig.~(\ref{fig4}) roughly resemble the temperature dependence of the purity,
		which is found using eq.~(\ref{prk}) with appropriate values of wave vectors
		$\bm{k}$, meaning that the dependence in the shown temperature range is 
		predominantly exponential decay. At high temperatures 
		(for nanostructures, meaning far outside the $20$ K range of Fig.~(\ref{fig4})),
		the decay is dominated by terms proportional to $1/T^3$. The fitted dependence of Negativity
		on temperature is presented at the end of Sec.~(\ref{secVC}), since the
		dependence on temperature is convoluted with the dependence on environment
		size and cannot be considered separately.
		
		Note that the trade-off temperature behavior, 
		which is characteristic for the build up of correlations in the studied system
		\cite{krzywda} and which results from the decrease of purity with temperature
		accompanied by an increase of the overall effect of the environment on the 
		qubit, is not present here, as only the purity of the system state is relevant
		for the generation of entanglement, as long as the system-environment 
		interaction is capable of entangling the two subsystems.
		Contrarily, this type of trade-off behavior has been reported for boson-boson system-environment ensembles \cite{hilt09}.
		
		\subsection{Dependence on environment size - pure initial state \label{secVB}}

In the case of a pure initial state of the environment (at zero temperature in the case
of the studied system, so there
are initially no phonons), the joint evolution of the system and the environment remains pure
and entanglement at any time can be evaluated
in a straightforward manner using the von Neumann entropy of one of the entangled
subsystems (such von Neumann entropy is the 
unique entanglement measure for pure states).
The measure is defined as
\begin{equation}
\label{von}
E(|\psi(t)\rangle)=-\frac{1}{\ln 2}\Tr\left(
\rho(t)\ln\rho(t)
\right),
\end{equation}
where $|\psi(t)\rangle$ is the pure system-environment state
and $\rho(t)=\Tr_E|\psi(t)\rangle\langle \psi(t)|$
is the density matrix of the qubit at time $t$ (obtained by tracing out the
environment). The entanglement measure in eq.~(\ref{von}) is
normalized to yield unity for maximally entangled states.
The same result would be obtained when tracing out the qubit degrees of freedom
instead of the environmental degrees of freedom, but the small dimensionality
of the qubit makes this way much more convenient.

Let us denote
the pure initial state of the environment as $|R_0\rangle$. Then
qubit-environment state at time $t$ is given by
\begin{equation}
|\psi(t)\rangle = \alpha|0\rangle \otimes|R_0\rangle+\beta e^{i\epsilon t/\hbar}|1\rangle \otimes\hat{u}(t)|R_0\rangle,
\end{equation}
and $\hat{u}(t)$ is given by eq.~(\ref{u}).
The density matrix of the qubit is now of the form
\begin{equation}
\label{rho}
\rho(t)=\left(
\begin{array}{cc}
|\alpha|^2&\alpha\beta^*e^{-i\epsilon t/\hbar}u^*(t)\\
\alpha^*\beta e^{i\epsilon t/\hbar} u(t)&|\beta|^2
\end{array}
\right),
\end{equation}
where $u(t)=\langle R_0|\hat{u}(t)|R_0\rangle$
and the absolute value of the function $u(t)$
constitutes the degree of coherence retained in the qubit system
at a given time (it is given by eq.~(\ref{srednia}) with $T=0$).

The entanglement measure of eq.~(\ref{von}) can be calculated using eq.~(\ref{rho})
which yields
\begin{eqnarray}
E(|\psi(t)\rangle)&=&-\frac{1}{\ln 2}
\left[
\frac{1+\sqrt{\Delta(t)}}{2}\ln\frac{1+\sqrt{\Delta(t)}}{2}\right.\\
\nonumber
&&+\left.
\frac{1-\sqrt{\Delta(t)}}{2}\ln\frac{1-\sqrt{\Delta(t)}}{2}
\right],
\end{eqnarray}
with $\Delta(t)=1-4|\alpha|^2|\beta|^2+|\alpha|^2|\beta|^2|u(t)|^2$.
Note that the von Neumann entropy during pure dephasing depends only
on the degree of coherence $|u(t)|$. This means that, if two qubits
lose the same amount of coherence, they must be entangled with the environment to
the same degree, regardless of the numer of boson modes which constitute the 
environment. Hence, for pure initial environmental states, the amount of entanglement 
generated during evolution does not depend on the size of the environment, as long as the degree of coherence at a given time does not depend on its size.

A similar analysis can be performed using Negativity as the measure of pure
state entanglement (Negativity does not converge to von Neumann entropy for pure
states contrarily to most entanglement measures). It is then fairly straightforward
to show that entanglement does not depend on the size of the environment, but only
on the degree of coherence $u(t)$, taking into account the fact that Negativity in the studied system
does not depend on the phase relations between different components of the density 
matrix $\hat{\sigma}$. What is not straightforward is obtaining the
explicit relation between Negativity and decoherence, and therefore the von Neumann
entropy was used in the analysis above.

\subsection{Dependence on environment size \label{secVC}}

\begin{figure}[th]
	\begin{center}
		\unitlength 1mm
		\begin{picture}(75,55)(5,5)
		\put(0,0){\resizebox{85mm}{!}{\includegraphics{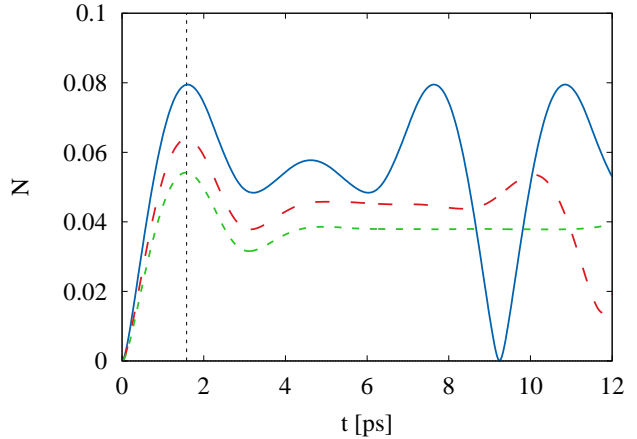}}}
		\end{picture}
	\end{center}
	\caption{\label{fig5} Entanglement evolution for three
		different numbers of boson modes ($n=6$ - blue solid line, $n=8$ - red dashed
		line, $n=10$ - green dotted line) at temperature $T=6$ K. Vertical line indicates the first time at which entanglement is maximized.}
\end{figure}

\begin{figure}[th]
	\begin{center}
		\unitlength 1mm
		\begin{picture}(75,55)(5,5)
		\put(0,0){\resizebox{85mm}{!}{\includegraphics{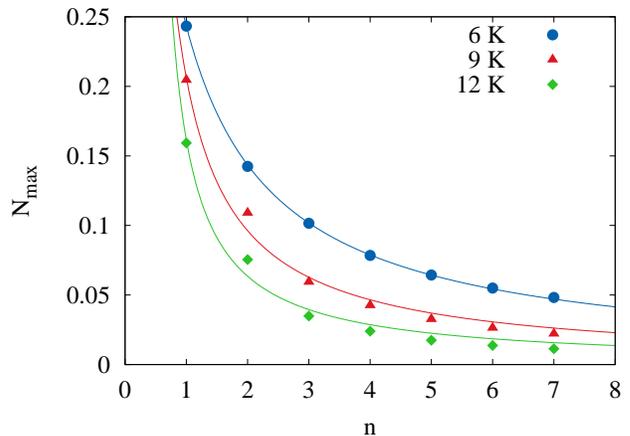}}}
		\end{picture}
	\end{center}
	\caption{\label{fig6} Maximal entanglement as a function of the number
		of boson modes for different temperatures. The points correspond to numerical
		data (blue dots - $6$ K, red triangles - $9$ K, green squares - $12$ K), while the
		lines depict the fitting function and are color-coded in the same way.}
\end{figure}

At finite temperatures (for non-pure initial environmental states)
the number of boson modes taken into account becomes very important.
In Fig.~(\ref{fig5}) the evolution of Negativity for the equal superposition
initial state of the qubit ($\alpha=\beta=1/\sqrt{2}$) is plotted at $T=6$ K
for different numbers of boson modes $n$. Note that the temperature is high enough,
so that the initial drop of qubit coherence is always the same as in the continuous 
case, while the platou is reproduced for some short time after the drop
up to small oscillations
(and this time is longer for larger $n$). This means that the three curves in 
Fig.~(\ref{fig5}) correspond to the same qubit decoherence curves at short times
(up to roughly $5$ ps here). Obviously, the amount of entanglement generated
during these decoherence processes is not the same, as both the maximum values and
the values at the plateau decrease with increasing number of boson modes.
As the temperature dependence, this is related to the purity of the whole system
during the evolution, but the temperature also affects the decoherence curves,
while the number of phonon modes (when they are chosen as outlined in Sec.~\ref{sec2})
does not.

The dependence of maximum Negativity as a function of the number of boson modes
taken into account is plotted in Fig.~(\ref{fig6}) with points 
for three different temperatures
(well above the threshold value). The decrease of Negativity with growing $n$ is 
rather steep for the temperatures shown and this steepness increases when the temperature
grows. This corresponds to the fact that at zero temperature, entanglement does not
depend on the size of the environment, while at infinite temperature
no entanglement between the qubit and the environment is generated at all 
\cite{roszak15a}. Furthermore, for any finite temperature entanglement approaches zero
with growing $n$ according to a function proportional to $1/n$, and for a continuous environment, no entanglement is generated
in the system. Although separability is reached more slowly at lower temperatures
it is reached nonetheless for large enough values of $n$ (technically, zero-Negativity
is only obtained for $n=\infty$, but for high enough $n$ the values of Negativity
will be so small that such entanglement will no longer be detectable and will have
practically no effect on the properties of the system).

Fitting of the curves displayed in Fig.~(\ref{fig4}) for temperatures
above the threshold temperature and points displayed in
Fig.~(\ref{fig6}) allows to find the dependence of maximum Negativity on temperature
and environment size.
The dependence on size exhibits good $\sim \frac{1}{n}$ behavior.
The temperature dependence, on the other hand, shows strong exponential decay 
for low temperatures and, while
increasing temperature, a $1/T^3$ dependence becomes dominant. 
Furthermore, the temperature and size dependencies are convoluted, so they cannot
be represented as a simple product of temperature-dependent and size-dependent functions.
A reasonable fit is obtained using the function
\begin{equation}
\label{fitting}
N_{\mathrm{max}}(n,T)\approx e^{-\alpha T}\frac{A}{T(n-BT+CT^2+D)},
\end{equation}
where the fitting parameters are given by $\alpha=0.0857$, $A=3.51$, 
$B=0.4674$, $C=0.01865$, and $D=2.57$.
The curves obtained using eq.~(\ref{fitting}) are also plotted on Fig.~(\ref{fig6}),
showing a very good $n$-dependence for different temperatures, especially for higher
numbers of boson modes ($n\ge 3$).

\section{Conclusion \label{sec6}}

We have studied the generation of entanglement quantified by Negativity
between a charge qubit and its bosonic environment during evolution
which leads to pure dephasing of the qubit. 
In particular, we studied and excitonic qubit confined in a quantum dot
in the presence of a super-Ohmic phonon bath, but the results could be easily extended
to other charge qubits undergoing similar decoherence processes.
The quantity of interest
was the dependence of the amount of generated entanglement on the size of the environment
(the number of boson modes taken into account) in the situation,
when the evolution of the qubit alone does not depend on environment size
(for short enough times and high enough temperatures, such a situation is 
easily obtained). We have found that although for pure states entanglement does
not depend on the system size (and the amount of generated entanglement for pure
dephasing is an explicit function of the degree of qubit coherence), for finite
temperatures Negativity is a decreasing function of environment size proportional to
$1/n$ and there is no entanglement generated
for a continous bosonic environment regardless of the temperature (as long as $T\neq 0$).

The temperature, which governs the initial mixedness of the environment, and consequently
the mixedness of the whole system throughout its unitary evolution, similarly
governs entanglement generated between the system and environment.
This means that for higher temperatures, the state of the whole system is more mixed,
so less entanglement is generated. 
In fact, for reasonably low temperatures, the temperature-dependence of the maximum
Negativity reached during the joint evolution is almost exponential and roughly resembles
the temperature-dependence of the purity.
This dependence is obviously monotonously decreasing
and does not display the trade-off resulting from the fact that the effect of the 
environment on the qubit is stronger at higher temperatures while purity is 
decreased.

\end{document}